\documentclass[english,aps,prl,notitlepage,twocolumn,superscriptaddress,groupedaddress]{revtex4}

\usepackage[T1]{fontenc}
\usepackage[latin9]{inputenc}
\usepackage{color}
\usepackage{amsmath}
\usepackage{amssymb}
\usepackage{graphicx}
\usepackage{amssymb}
\usepackage{epsfig}
\usepackage{wasysym}
\usepackage{epstopdf}
\usepackage{bm}
\usepackage{graphicx,epsfig}
\usepackage{mathrsfs}
\usepackage{dcolumn}
\usepackage{natbib}
\usepackage{CJK}
\usepackage{miniplot}
\usepackage{subfigure}
\makeatletter

\usepackage{amsmath}
\newcommand\ket[1]{\left| #1\right\rangle}
\newcommand\bra[1]{\left\langle #1\right|}
\newcommand\braket[2]{\left\langle #1| #2\right\rangle }
\allowdisplaybreaks[2]
\makeatother

\usepackage{babel}
\begin{document}
\global\long\def\ket#1{\left| #1\right\rangle }

\global\long\def\bra#1{\left\langle #1\right|}

\global\long\def\braket#1#2{\left\langle #1| #2\right\rangle }

\title{Topological phases of quantized light }

\author{Han Cai}
\affiliation{Interdisciplinary Center for Quantum Information and State Key Laboratory of Modern Optical Instrumentation, Zhejiang Province Key Laboratory of Quantum Technology and Device and Department of Physics, Zhejiang University, Hangzhou 310027, China}

\author{Da-Wei Wang}
\email{dwwang@zju.edu.cn}
\affiliation{Interdisciplinary Center for Quantum Information and State Key Laboratory of Modern Optical Instrumentation, Zhejiang Province Key Laboratory of Quantum Technology and Device and Department of Physics, Zhejiang University, Hangzhou 310027, China}
\affiliation{CAS Center for Excellence in Topological Quantum Computation, University of Chinese Academy of Sciences, Beijing 100190, China}

\date{\today}

\begin{abstract}
Topological photonics is an emerging research area that focuses on the topological states of classical light. Here we reveal the topological phases that are intrinsic to the quantum nature of light, i.e., solely related to the quantized Fock states and the inhomogeneous coupling between them.~The Hamiltonian of two cavities coupled with a two-level atom is an intrinsic one-dimensional Su-Schriefer-Heeger model of Fock states. By adding another cavity, the Fock-state lattice is extended to two dimensions with a honeycomb structure, where the strain due to the inhomogeneous coupling strengths of the annihilation operator induces a Lifshitz topological phase transition between a semimetal and three band insulators within the lattice. In the semimetallic phase, the strain is equivalent to a pseudomagnetic field, which results in the quantization of the Landau levels and the valley Hall effect. We further construct an inhomogeneous Fock-state Haldane model where the topological phases can be characterized by the topological markers. With $d$ cavities being coupled to the atom, the lattice is extended to $d-1$ dimensions with no upper limit.~This study demonstrates a fundamental distinction between the topological phases of bosons and fermions and provides a novel platform for studying topological physics in dimensions higher than three.
\end{abstract}

\maketitle

Topological phases of matter have been extensively investigated not only in electrons \cite{Klitzing1980,Haldane1988,Kane2005,Bernevig2006}, but also in neutral atoms \cite{goldman2010realistic,Jotzu2014},
photons \cite{Lu2014,khanikaev2017two}
and phonons \cite{yang2015topological,serra2018observation}.
However, regarding whether the topological phases are quantum or classical, 
there is a fundamental difference between electrons and photons (and similarly phonons).~While the topological phases of electrons are intrinsically quantum, i.e., based on the Schr\"odinger equation and fermionic statistics of electrons, the topological phases of light originating from the analogy between the Maxwell and Schr\"odinger equations can be explained in the framework of classical optics \cite{Lu2014, khanikaev2017two, ozawa2019topological}. Although in lattices of resonators \cite{Hafezi2011} a quantized field formulation of light is used to facilitate the calculation of the chiral edge modes in parallel with those of electrons, the topological phases have no quantum signature and can be demonstrated with classical light.~A natural question is whether the second quantization of light embeds new topological phases that are fundamentally distinct from those classical ones.
Such topological phases of quantized light can bring together two relatively unrelated areas, the quantum electrodynamics and the topological matter, and provide a new perspective on the relations between different topological phases in condensed matter physics.

In search of the topological phases of quantized light, it is surprising to notice that only a few examples in quantum optics require the field quantization, including the black-body radiation, the Lamb shift \cite{Lamb1947}, and the Casimir effect \cite{Casimir1948}. The former reveals the quantized eigenstates of light, named as the Fock states and denoted by $\ket{m}$ with $m$ being the number of photons in the states. The latter two are resulted from the quantum fluctuations of the vacuum state $\ket{0}$. The quantized Fock states have profound consequences in the atom-photon interactions, such as the collapse and revival of the Rabi oscillations  \cite{Eberly1980,Rempe1987,Brune1996} when a two-level atom is resonantly coupled to a coherent field, i.e., in the Jaynes-Cummings (JC) model \cite{jaynes1963comparison}. The collapse and revival are due to the quantum interference between the Rabi oscillations of the atom coupled to different Fock states $\ket{m}$, which have discrete Rabi frequencies proportional to $\sqrt{m}$. Another famous example that follows the $\sqrt{m}$-scaling are the Landau levels of electrons near the Dirac cones of a graphene in a magnetic field \cite{neto2009electronic,goerbig2011electronic}. In this paper, among other interesting connections between the JC model and the topological phases in condensed matter physics, we reveal the surprising relation between the $\sqrt{m}$-scaling of the Rabi frequencies and the Landau levels through a lattice composed by Fock states, coined the Fock-state lattice (FSL) \cite{WangCaiLiuEtAl2016}. 

\begin{figure}
\includegraphics[width=0.9\columnwidth]{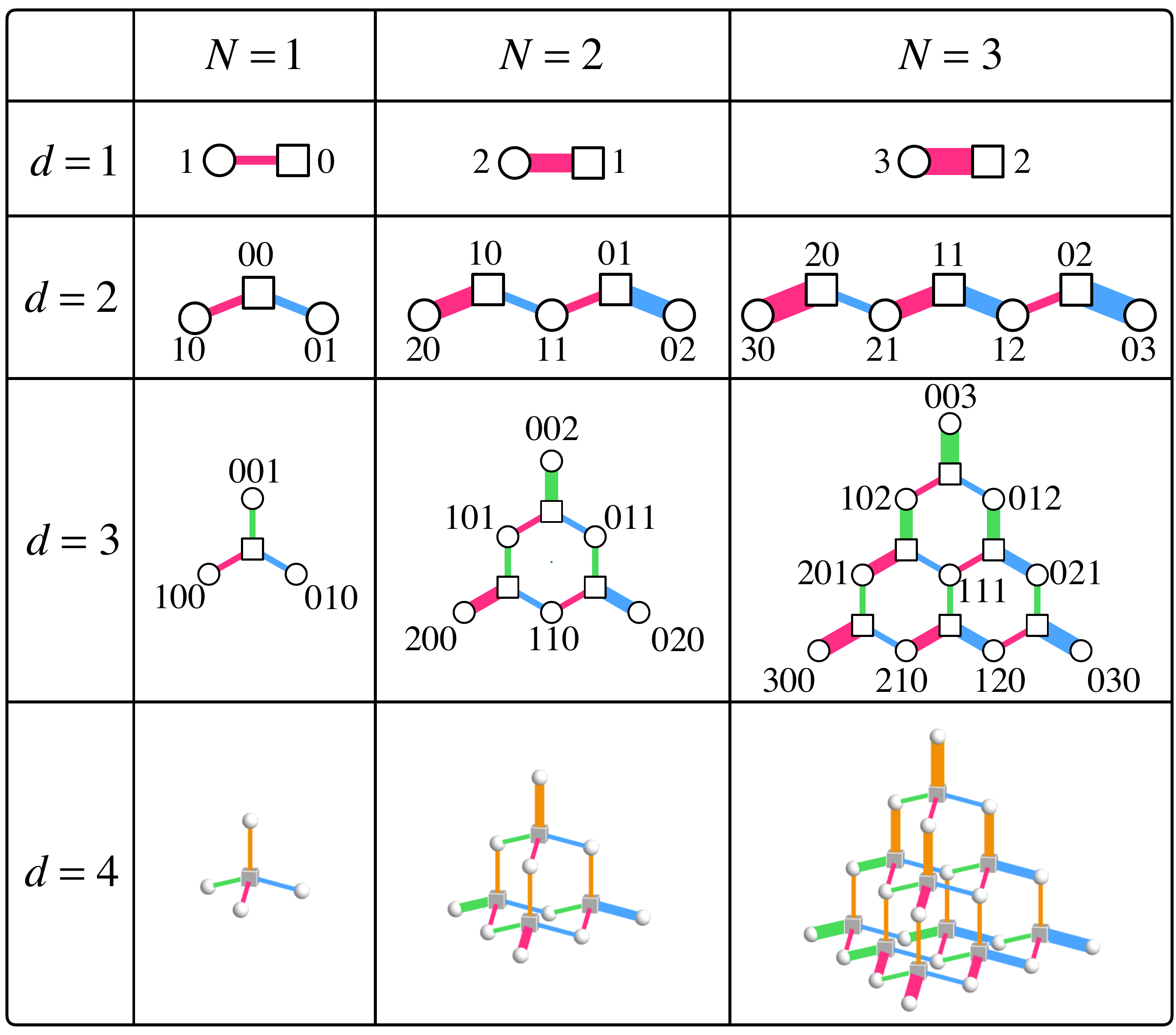}
\caption{Fock-state lattices in $d-1$ dimensions of the Hamiltonian in Eq.~(\ref{h}) with the total excitation number $N$. The squares/circles denote the states $\ket{\uparrow/\downarrow, n_1, n_2, ..., n_d}$ in the sublattices characterized by the $\ket{\uparrow}/\ket{\downarrow}$ atomic states. The numbers labeling the lattice sites are the photon numbers $n_1n_2...n_d$ in the corresponding states. For clarity we only label the photon numbers in the $\ket{\downarrow}$-sublattice for $d=3$ and hide all photon numbers for $d=4$. The widths of the lines connecting neighboring sites are proportional to the magnitudes of the coupling strengths between them.}
\label{fig0}
\end{figure}

Before we sketch the basic structure of the FSL, we emphasize that bosonic Hamiltonians have a much larger Hilbert space than the fermionic ones with the same number of modes, since each fermionic mode can only host one fermion due to the Pauli exclusion principle, while arbitrary number of bosons can stay in the same mode. This allows quantized bosonic states to form lattices with only a few number of modes. The Fock states of $d$ modes of photons are $\ket{n_1,n_2,...,n_d}$, where $n_j=0, 1, 2, ...$ is the photon number in the $j$th mode. Each mode offers an independent degree of freedom. We introduce the FSL with the Hamiltonian of a multi-mode JC model  ($\hbar=1$),
\begin{equation}
H=\frac{g}{\sqrt{d}}\sum_{j=1}^{d} (a_j^\dagger\sigma^-+\sigma^+a_j),
\label{h}
\end{equation}
where $\sigma^-=\ket{\downarrow}\bra{\uparrow}$ and $\sigma^+=\ket{\uparrow}\bra{\downarrow}$ are the lowering and raising operators of the two atomic states $\ket \uparrow$ and $\ket \downarrow$, $a_j$ and $a_j^\dagger$ are the annihiliation and creation operators of the $j$th mode, and $g/\sqrt{d}$ is the coupling strength between the photons and the atom.
This Hamiltonian conserves the total number of excitation $N=\sum_j a_j^\dagger a_j +(\sigma_z+1)/2$ where $\sigma_z=\ket{\uparrow}\bra{\uparrow}-\ket{\downarrow}\bra{\downarrow}$ is the $z$-component of the Pauli matrices of the atom.
We have two ways to look into the Hamiltonian in Eq.~(\ref{h}).
Each state $\ket{\uparrow, n_1, n_2, ..., n_d}$ is coupled to $d$ neighbors $\ket{\downarrow, n_1, n_2, ..., n_j+1,...,n_d}$ (where $j=1,2, ...,d$) with coupling strengths proportional to $\sqrt{n_j+1}$, forming
a bipartite (corresponding to the two states of the atom) FSL with site-dependent coupling strengths in $d-1$ dimensions (see Fig.~\ref{fig0}). From another perspective by combining the $a$ modes to form a collective mode $b=\sum_j a_j/\sqrt{d}$, the Hamiltonian becomes the single-mode JC model, which is analytically solvable. 
Combination of these two pictures enables us to study the topological phases of the FSL.

Before laying out the details, we first highlight a couple of distinctive features of the FSL.
They are lattices of quantum states instead of modes and have natural edges based on the fact that the photon numbers in Fock states have a lower limit zero, i.e., the existence of the vacuum state.
An advantage of the FSL is that their dimensions have no upper limit, providing a unique platform to investigate topological phases in dimensions higher than three.~However, we must take special care of the coupling strengths, which vary 
locally depending on the photon numbers in the Fock states. Here we show that for one-dimensional (1D) FSL with $d=2$, the variation of the coupling strengths results in the soliton state between two different topological phases of the Su-Schriefer-Heeger (SSH) model \cite{su1979solitons,heeger1988solitons}. 
In two dimensions with $d=3$, the variation of the coupling strengths is equivalent to a strain field in the honeycomb lattice, which leads to a Lifshitz topological phase transition between a semimetal and three band insulators within the lattice \cite{goerbig2011electronic}, as well as a strain-induced pseudomagnetic field \cite{Suzuura2002,pereira2009tight} in the semimetallic phase. The pseudomagnetic field results in the quantized Landau levels and provides the basis to observe the valley Hall effect \cite{xiao2007valley,yao2008valley,xiao2012coupled} and construct the Fock-state Haldane model \cite{Haldane1988}, where the topological phases are characterized by topological markers \cite{kitaev2006anyons,bianco2011mapping}. The FSL can be extended to higher dimensions to study the topological phases that are not achievable in real space \cite{zhang2001four, zilberberg2018photonic, lohse2018exploring}. It also provides a solution to design finite lattices with exactly quantized energy levels \cite{Guinea2010,Yang2017}. \\

\noindent\textbf{Results}

\begin{figure}
\includegraphics[width=0.9\columnwidth]{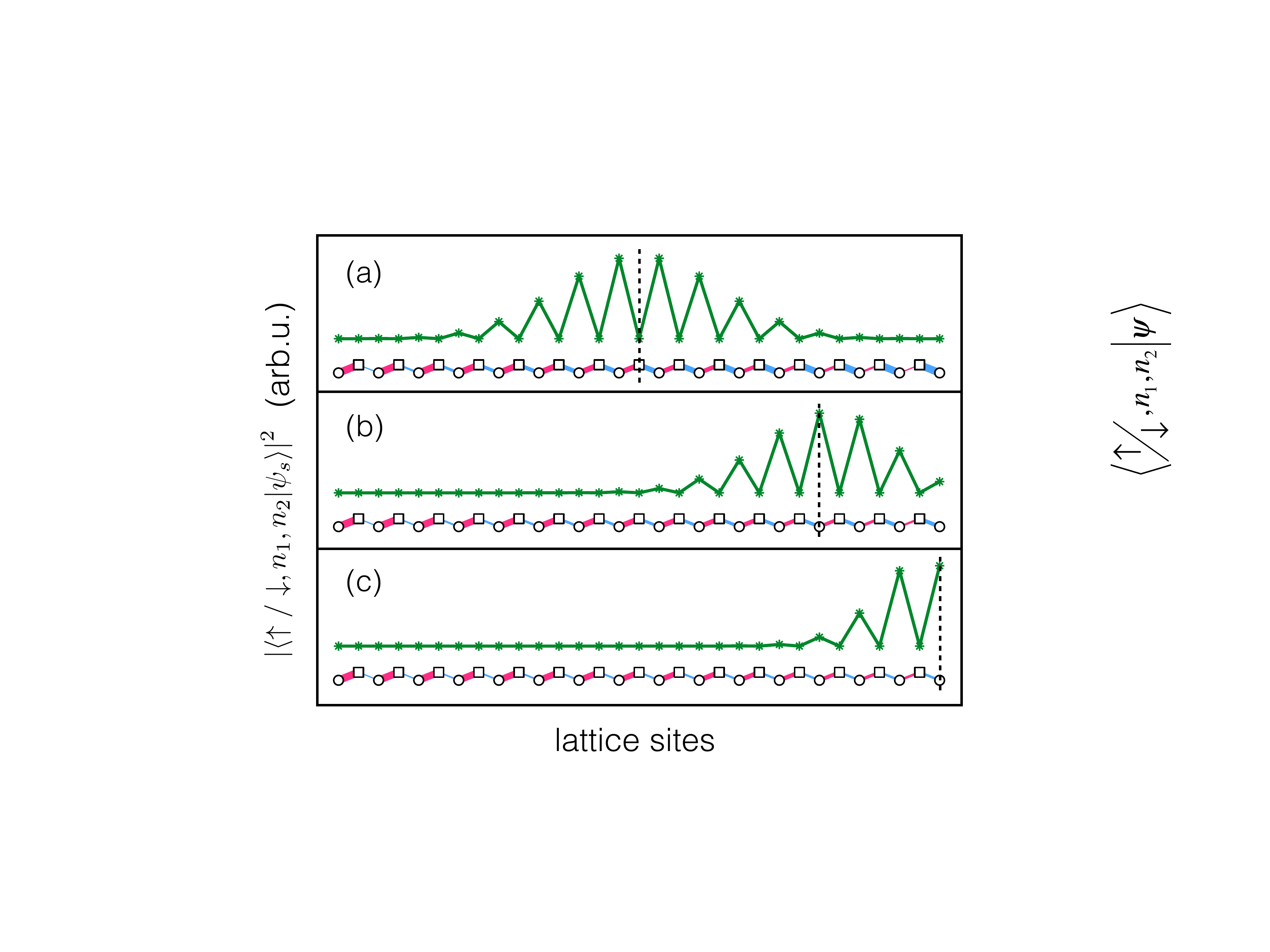}
\caption{The probability distribution of the soliton state $\ket{\psi_s}$ in the 1D Fock-state SSH model. The lattices are plotted in the same way as in Fig.~\ref{fig0} with $N=15$. The probability distribution of $\ket{\psi_s}$ is plotted above the corresponding lattice sites. The ratio $u_1/u_2=1$ (a), 2 (b) and 4 (c). The neighboring probabilities are connected by straight lines to guide the eyes. The vertical dashed lines label the boundary between two topological phases of the SSH model. }
\label{fig1}
\end{figure}

\begin{figure*}
\includegraphics[width=1.8\columnwidth]{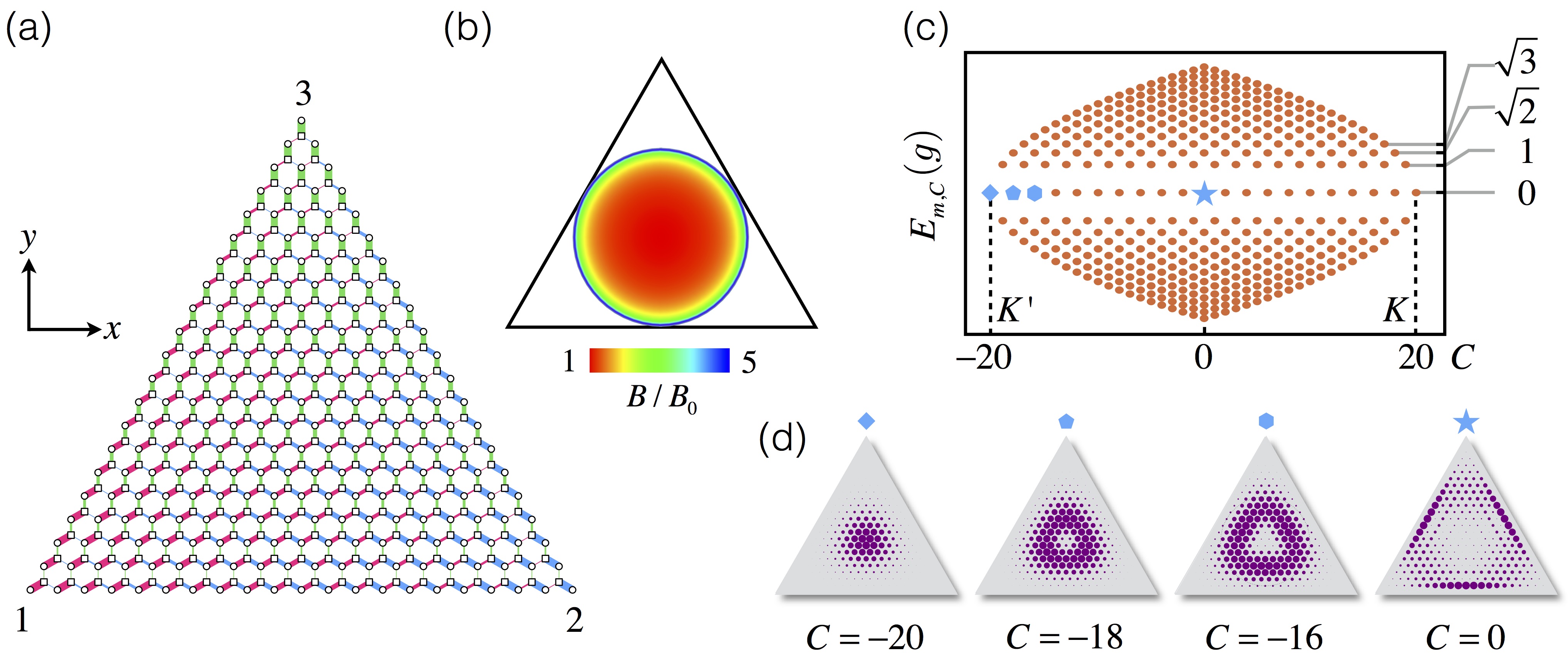}\caption{Two-dimensional Fock-state lattice with effective pseudomagnetic field and Landau levels.
(a) The Fock-state lattice of the Hamiltonian $H_2$ in Eq.~(\ref{2d}) for $N=20$. The three numbers 1, 2, 3 denote the states at the vertices with all $N$ photons in modes $a_1$, $a_2$ and $a_3$. The coupling strengths $t_1, t_2$ and $t_3$ are drawn with red, blue and green lines with widths proportional to the strengths. (b) The distribution of the effective pseudomagnetic field due to the variation of the coupling strengths within the incircle, evaluated from Eq.~(\ref{Br}). Outside of the incircle the strain induces a bandgap. (c) The band structure of the generalized Landau levels with eigenenergies $E^\pm_{m,C}=\pm \sqrt{m}g$ for the eigenstates $\ket{\psi^\pm_{m,C}}$. (d) The wavefunctions of the eigenstates in the zeroth Landau level $\ket{\psi_{0,C}}$ for $C=-20, -18, -16$ and 0, labelled with diamond, pentagon, hexagon and star in (c).}
\label{fig2}
\end{figure*}

\noindent\textbf{1D Fock-state SSH model.} We first show the relation between the SSH model and the 1D FSL with the Hamiltonian,
\begin{equation}
H_1=g\sigma^+(u_1 a_1+u_2 a_2) +h.c.,
\label{ssh}
\end{equation}
where $u_1$ and $u_2$ are real positive numbers satisfying $u_1^2+u_2^2=1$. In the insets of Fig.~\ref{fig1} (a)-(c),
we illustrate the Hamiltonian with the FSL in the subspace with $N=15$ in the basis
of $\left|\downarrow/\uparrow, n_1, n_2\right\rangle$, including $2N+1=30$ states, for different values of $u_1/u_2$. 
The connection between this lattice and the topological SSH model 
is endorsed by the variation of the coupling
strengths due to the property of the annihilation operator,
$a|n\rangle=\sqrt{n}|n-1\rangle$. For $u_1=u_2=1/\sqrt{2}$,
the lattice is equally divided into two parts. On the left side, the coupling strengths of
$a_1$ are larger than those of $a_2$, in contrary to their relation on the right side.
Accordingly, these two parts are in two different topological phases of the SSH model, which
is evident from the soliton state at the boundary, as shown in Fig.~\ref{fig1} (a)-(c). We can tune $u_1$ and $u_2$ to move the soliton state, which is always located at the boundary 
satisfying $u_1\sqrt{n_1}=u_2\sqrt{n_2}$ (see Methods).
When $u_1>\sqrt{N}u_2$ (or $u_2>\sqrt{N}u_1$), there is only one topological phase and
the soliton state is on the edges of the lattice.

The eigenenergies and eigenstates of Eq.~(\ref{ssh}) are analytically obtained by
recombining $a_1$ and $a_2$ to form a bright mode $b_1=u_1 a_1+u_2 a_2$
and a dark mode $b_2=u_2 a_1-u_1a_2$. Only the bright mode is coupled with the
atom. The corresponding eigenstates are $|\psi_{m}^\pm\rangle=(\left|\downarrow,m,N-m\right\rangle_b\pm\left|\uparrow,m-1,N-m\right\rangle_b)/\sqrt{2}$
where $m=1,2,...,N$ in $\ket{...}_b$ is the photon number in $b_1$ mode. 
The eigenstate with $m=0$ is
the soliton state $|\psi_s\rangle=\left|\downarrow,0,N\right\rangle_b$, which has zero energy and only occupies 
the $\left|\downarrow\right\rangle$-sublattice.

~\\
\noindent\textbf{Effective strain, pseudomagnetic field and Landau levels in the 2D FSL.} The lattice is extended to two dimensions by adding a third cavity mode,
\begin{equation}
H_2=\frac{g}{\sqrt{3}}\sigma^+(a_1+a_2+a_3)+h.c..
\label{2d}
\end{equation}
The Fock states $\ket{\uparrow/\downarrow, n_1, n_2, n_3}$ form a honeycomb lattice with a triangular boundary on which one of the cavity modes is in the vacuum state, as shown in Fig.~\ref{fig2} (a). Only one cavity has photons at the three vertices, which are labelled with the corresponding cavity numbers.~Again the coupling strengths are inhomogeneous,
which introduces an effective strain in the lattice. 
We first notice that in the center of the lattice the strain is relatively small while approaching the vertices the strain becomes drastic.
When the strain is small such that \cite{pereira2009tight},
\begin{equation}
|t_1-t_2|< t_3 < |t_1+t_2|,
\label{tb}
\end{equation} 
with $t_j=g\sqrt{n_j}/\sqrt{3}$ being the coupling strength of the mode $a_j$, the strain field is equivalent to a pseudomagnetic field leading to the quantized Landau levels \cite{Suzuura2002,pereira2009tight,Guinea2010}, which have been experimentally implemented in graphene \cite{Levy2010}.
The lattice sites that satisfy Eq.~(\ref{tb}) are in the incircle of the FSL, i.e., where (see Fig.~\ref{fig2} (b) and Methods) 
\begin{equation}
n_1^2+n_2^2+n_3^2<\frac{N^2}{2}.
\label{innernumbers}
\end{equation}
Beyond the incircle the strain is so large that a bandgap opens
and we cannot regard the strain as a simple pseudomagnetic field. 
A Lifshitz topological phase transition between a strained semimetal and a band insulator \cite{goerbig2011electronic} happens on the incircle. 

We first evaluate the strength of the pseudomagnetic field near the center of the FSL. This can be achieved by comparing the eigenenergies of Eq.~(\ref{2d}) and the Landau levels in a real graphene, which is characterized by $\pm\sqrt{mB}$-scaling near the Dirac cone, with $B$ being the
strength of the magnetic field, $m$ being the index of the
Landau levels and $\pm$ for the conduction and valence bands \cite{goerbig2011electronic}. 
The eigenenergies of the Hamiltonian $H_2$
can be obtained by recombining the cavity modes to form a collective bright mode,
$b_0=(a_1+a_2+a_3)/\sqrt{3}$. The JC model of $b_0$ mode coupling with the atom has
eigenenergies $\pm \sqrt{m} g$ with $m=\langle b_0^\dagger b_0\rangle$, 
i.e., in accord with the scaling of the Landau levels in a graphene, with the effective cyclotron frequency $g$.
By recalling the explicit form of the energies of Landau levels in a graphene \cite{goerbig2011electronic} and comparing it with the eigenenergies
of $H_2$, we obtain
\begin{equation}
\pm\sqrt{m}g=\pm \sqrt{2m} \frac{3t_h q}{2l_B}, 
\label{ll}
\end{equation}
where $t_h$  is the hopping coefficient and $q$ is the lattice constant, and the magnetic length $l_B=\sqrt{\hbar/eB}$ with $e$ being the electric charge.

At the center of the honeycomb FSL where $\langle a_j^\dagger a_j\rangle \approx N/3$ for $j=1,2,3$, 
the coupling strengths are $t_1=t_2=t_3=t_h\equiv\sqrt{N}g/3$,
which can be regarded as
the unstrained background hopping coefficient. The pseudomagnetic field is built upon
the deviation of the coupling strengths from $t_h$ due to the variation of the photon numbers.
Substituting $t_h$ in Eq.~(\ref{ll}), we obtain
\begin{equation}
\frac{l_B}{q}=\sqrt{\frac{N}{2}},
\label{ml}
\end{equation}
which is the only relevant quantity to measure the strength of the pseudomagnetic field
since both $q$ and $l_B$ are fictious in the FSL. The strength of the corresponding pseudomagnetic field is
\begin{equation}
B_0=\frac{2\hbar}{Neq^2}.
\end{equation}
The fictious electric charge $e$ in $B_0$ is only an analogous quantity for the convenience of comparison with electrons. All observables in the lattice are independent of $e$.
However, to have a general idea of the strength of $B_0$, we take the lattice constant $q=0.14$ nm of graphene and obtain $B_0=6.5\times 10^4/N$ Tesla.
For a typical $N=20$, $B_0$ is 10 times larger than those achievable in graphene \cite{Levy2010}.

The pseudomagnetic field can only be regarded as approximately uniform near the center of the lattice.
The explicit distribution of the pseudomagnetic field is obtained through the valley Hall response (see Eq.~(\ref{Br})), or directly from the strain-induced motion of the Dirac cones (see Methods). 
Interestingly, despite the complications of the nonuniform pseudomagnetic field and the the topological phase transition on the inner cirlce, all the eigenstates in the 2D FSL are grouped in quantized energy levels with the $\pm\sqrt{m}$-scaling. In the following, we regard these
levels as generalized Landau levels of the FSL.

The degeneracy of the eigenstates in the $m$th Landau level is $N-m+1$.
To distinguish these states, we introduce the bosonic chirality operator,
\begin{equation}
C=b_+^\dagger b_+-b_-^\dagger b_-,
\label{c}
\end{equation}
where $b_\pm=\sum_{j=1}^3 a_j\exp{(\mp i 2j\pi/3)}/\sqrt{3}$ are the annihilation operators of the two dark modes. $C$ characterizes the momentum (or angular momentum considering the ring configuration of the three cavities) carried by the photons in the eigenstates. 
This quantity is an extension of the spin chirality \cite{Wen1989} (see Methods). 
In graphene, the $K$ and $K^\prime$ points correspond to
the two maximum momenta in the Brillouin zone \cite{neto2009electronic}. In the finite FSL the points with
 $C=N$ and $C=-N$ are the counterparts of the $K$ and $K^\prime$
points. The band structure of the 2D FSL is
shown in Fig.~\ref{fig2} (c).

The eigenstates in the $m$th Landau level are $|\psi^\pm_{m,C}\rangle=(\left|\downarrow, m, m_+, m_-\right\rangle_b \pm \left|\uparrow, m-1, m_+, m_-\right\rangle_b)/\sqrt{2}$, where $m_+$ and $m_-$ are the photon numbers
in the two dark modes. The $N+1$ eigenstates in the zeroth Landau level are solely composed by the $\left|\downarrow\right\rangle$-sublattice
states, $|\psi_{0,C}\rangle=\left|\downarrow, 0, m_+, m_-\right\rangle_b$, which are the counterparts of the soliton state in the 1D FSL.
We recall that in graphene the electrons in the zeroth Landau level of a real magnetic field only occupy
one sublattice at $K$ point and the other sublattice at $K^\prime$ point \cite{goerbig2011electronic}.
When the direction of the magnetic field is reversed, the zeroth-Landau-level occupations of the two sublattices at the $K$ and $K^\prime$ points are also reversed.
Since the strain-induced pseudomagnetic field has opposite signs at $K$ and $K^\prime$ points,
the states in the zeroth Landau level of the FSL only occupy the $\left | \downarrow\right\rangle$-sublattice
at both $K$ and $K^\prime$ points \cite{Gomes2012,wen2019acoustic}.

The wavefunctions of the eigenstates can be analytically obtained by making expansion
in the Fock states of $a$ modes. In Fig.~\ref{fig2} (d), we draw serveral eigenstates in
the zeroth Landau level. Near the $K^\prime$ point for $C=-20, -18$ and $-16$, the eigen wavefunctions resemble the ones 
in the zeroth Landau level of a real magnetic field with the symmetric gauge, but with a smaller localization length (see Methods). 
When $C$ decreases, the eigenstate approaches to the incircle of the triangular boundary,
as shown by $|\psi_{0, 0}\rangle$ in Fig.~\ref{fig2} (d) (see more wavefunctions in Methods).

\begin{figure}[t]
\includegraphics[width=0.95\columnwidth]{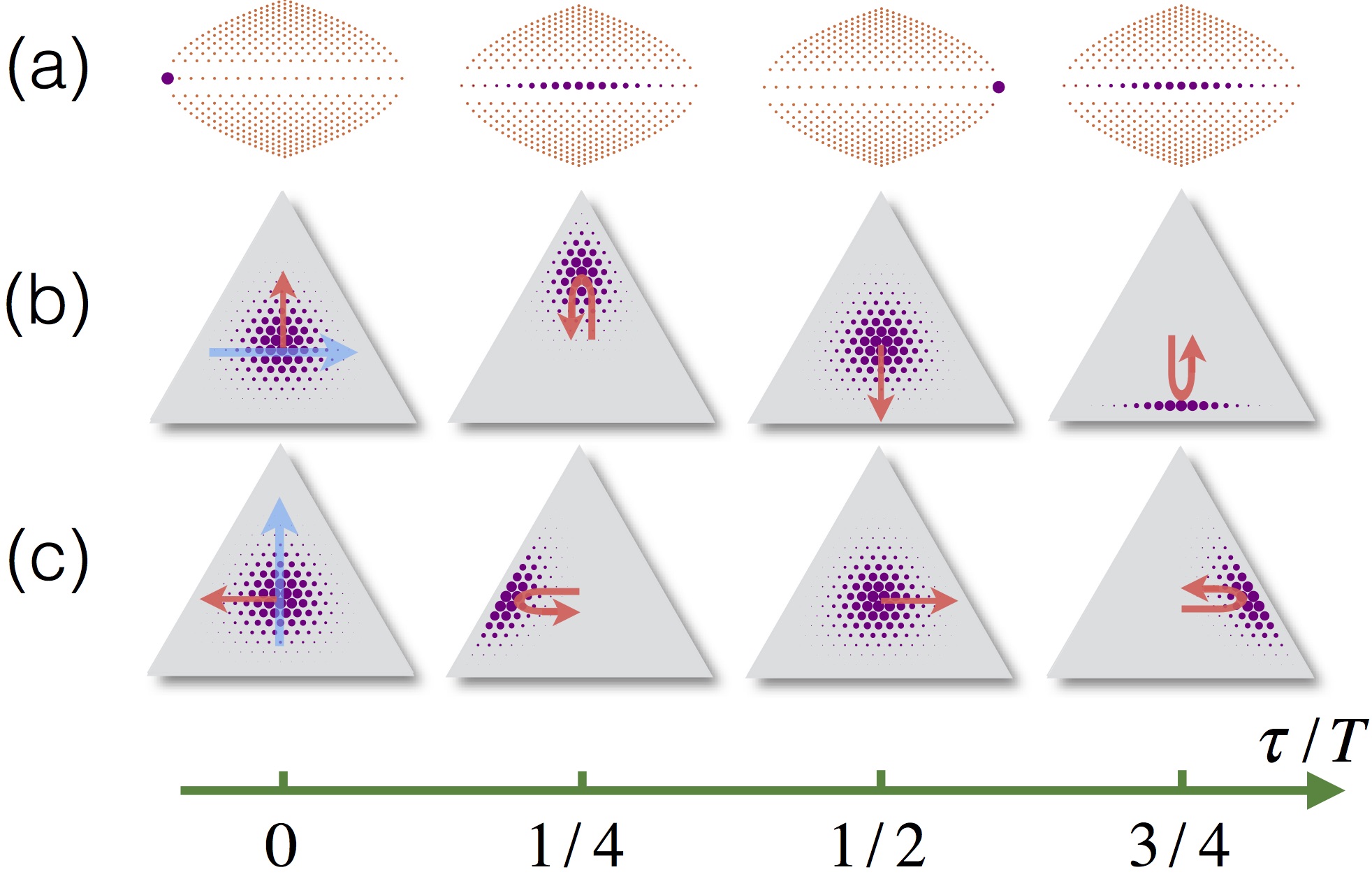}\caption{The Bloch oscillation and the valley Hall effect in the zeroth Landau level. (a) The evolution of the wavefunctions in the Landau levels for a small force with $\delta=0.01g$  (independent of the direction of the force). The total excitation number $N=20$. (b) and (c) show the dynamics of the wavefunctions in the FSL with forces in the directions of the blue arrows. The red arrows show the directions of the velocities at $\tau=0, T$. The U-turn arrows show the velocity change before and after $\tau=T/4, 3T/4$. The radii of the purple solid circles are proportional to the probabilities in the corresponding states.}
\label{fig5}
\end{figure}

~\\
\textbf{The valley Hall effect.} To directly see the opposite signs of the pseudomagnetic field, we need to introduce an effective electric field in the lattice and observe the Hall response of
states at $K$ and $K^\prime$ points. A static electric field induces a potential linear to the position of an electron.
In the FSL, such a linear potential can be introduced by the frequency difference between the cavity modes, e.g.,
\begin{equation}
H_3=H_2+\delta (a_1^\dagger a_1-a_2^\dagger a_2),
\label{h3}
\end{equation}
where $\delta$ is the detuning between $a_1$ and $a_2$ modes.
The direction of the effective force due to this potential is along the horizontal arrow in Fig.~\ref{fig5} (b). 

We prepare an initial state in the zeroth Landau level at the $K^\prime$ valley, $\ket{\psi(0)}=\ket{\psi_{0,-N}}$, and show its dynamical evolution with Hamiltonian $H_3$ by taking snapshots of the wavefunction at different times in Fig.~\ref{fig5}, where the distributions of the states in both the energy bands and the FSL are plotted. The electric field is small $\delta \ll g$ such that the Landau-Zener tunneling is negligible and the state stays in the zeroth Landau level. Driven by the effective electric field, the state moves from $K^\prime$ to $K$ (at time $\tau=T/2$ where $T=\sqrt{3}\pi/\delta$) and then returns to $K^\prime$ point, as shown in Fig.~\ref{fig5} (a), independent of the direction of the force. This is the Bloch oscillation in the zeroth Landau level. During this process, the most interesting feature of the valley Hall effect is demonstrated by the propagation of the wavefunction perpendicular to the direction of the force \cite{xiao2007valley}. In Fig.~\ref{fig5} (b) for a rightward force, the wavefunction moves upward at the $K^\prime$ point (when $\tau=0$) and downward at the $K$ point (when $\tau=T/2$), which is unambiguous evidence that the pseudomagnetic fields at $K$ and $K^\prime$ points have opposite signs. This effect can also be demonstrated with forces in any other directions, e.g., upward as shown in Fig.~\ref{fig5} (c) with the following force term in the Hamiltonian, $\delta(a_1^\dagger a_1+a_2^\dagger a_2-2a_3^\dagger a_3)/\sqrt{3}$.
The Landau-Zener tunneling appears when the potential difference between neighboring lattice sites $\delta$ is comparable or larger than the bandgap $g$ (see Methods).

We can calculate the drift velocity in the limit of small electric field when $\delta\ll g$ at the $K^\prime$ point through the standard formula \cite{ezawa2008quantum}, e.g., for a horizontal force as shown in Fig.~\ref{fig5} (b),
\begin{equation}
v_D=\frac{\mathscr{E}}{B_0}=\frac{Nq\delta}{\sqrt{3}},
\label{vd}
\end{equation}
where $\mathscr{E}=2\hbar\delta/\sqrt{3}qe$ is the strength of the effective electric field. On the other hand, from an independent approach (see Methods) the drifted center of the wavepacket follows a sinusoidal oscillation with amplitude $R= Nq/2$ (the radius of the incircle of the triangular boundary),
\begin{equation}
y(\tau)=R\sin{\frac{2\pi \tau}{T}},
\label{yt}
\end{equation}
where we have set the center of the lattice as the zero point. We obtain the velocity,
\begin{equation}
v_y(\tau)\equiv\frac{dy(\tau)}{dt}=v_D\cos{\frac{2\pi \tau}{T}}.
\label{vyt}
\end{equation}
Obviously, at $\tau=0$ it coincides with the drift velocity obtained from Eq.~(\ref{vd}), $v_y(0)=v_D$. At $\tau=T/2$, the wavepacket arrives at the $K$ point and $v_y(T/2)=-v_D$.

\begin{figure*}
\includegraphics[width=1.6\columnwidth]{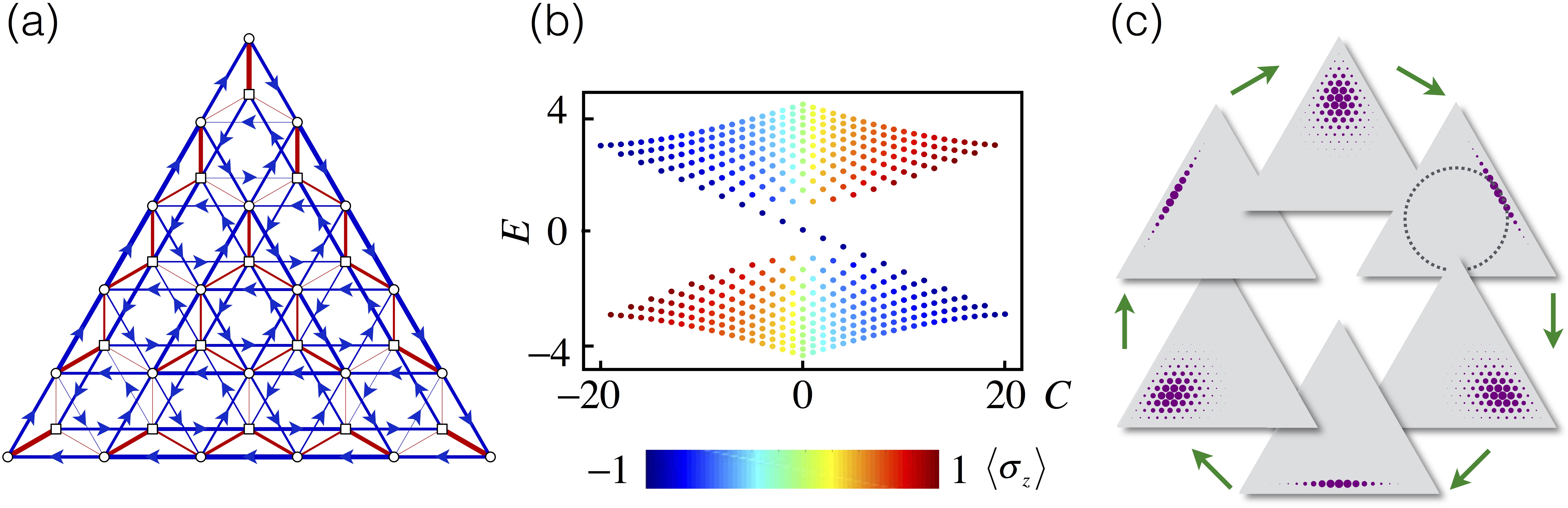}\caption{The Haldane model in the Fock-state lattice. (a) The coupling strengths of the Hamiltonian $H_4$ in Eq.~(\ref{h3}) with $N=5$. The nearest-neighbor couplings are denoted with red lines and the next-nearest-neighbor couplings are denoted by blue lines with arrows for the transition directions attached with a phase factor $i$. The linewidths are proportional to the coupling strengths. (b) The band structure of the Haldane model with $N=20$, $g=1$ and $\kappa=0.3$ in $H_4$. The color denotes the polarization of the eigenstates in $\ket{\uparrow}$ (red) and $\ket{\downarrow}$ (blue) components. (c) The dynamic evolution of a wavepacket of the edge states for $N=20$, starting from the one with an incircle denoting the trace of the weight (expectation value of the position) of this wavepacket during the evolution. The red arrows show the direction of time, sequentially at $\tau=nT_w/6$, where $T_w=2\pi/\kappa$ and $n=0,1,2,3,4,5$. }
\label{fig3}
\end{figure*}

Eqs.~(\ref{yt}) and (\ref{vyt}) also enable us to evaluate the strength of the pseudomagnetic field $B$ away from the center of the lattice through $B(y)=\mathscr{E}/v_y(y)$. Due to the rotational symmetry of the Hall response in this lattice, we obtain from Eqs.~(\ref{yt}) and (\ref{vyt}),
\begin{equation}
B^\pm(r)=\mp \frac{B_0}{\sqrt{1-r^2/R^2}},
\label{Br}
\end{equation}
where $r=\sqrt{x^2+y^2}$ is the distance to the center of the lattice, and $B^+(r)$ and $B^-(r)$ are for $K$ and $K^\prime$ valleys, respectively. The distribution of $B^-(r)$ is plotted in Fig.~\ref{fig1} (b) and the result is also consistent with a calculation based on the strain-induced shift of the Dirac cones (see Methods). In the $K^\prime$ valley, the total number of the magnetic flux quanta ($\Phi_0=2\pi\hbar/e$) in the incircle of the FSL is
${\int_0^R 2\pi rB^-(r) dr}/{\Phi_0}={N}/{2}$
which means that $N/2$ states can be hosted in the $K^\prime$ valley \cite{goerbig2011electronic}. On the other hand, there are $N+1$ eigenstates in the zeroth Landau level and half of them belong to the $K^\prime$ valley, which is consistent with the above result from the total magnetic flux.

~\\ \textbf{The Haldane model in the 2D FSL.} Although the 1D FSL is a topological SSH model, the 2D FSL
has a topologically trivial Chern number, evident from the absence of gapless edge states. 
However, by introducing additional terms in the Hamiltonian, we can construct a Haldane model,
\begin{equation}
H_4=H_2+\kappa \sigma_z C/2
\label{h4}
\end{equation}
where $\kappa$ is a coupling constant and
the bosonic chirality operator $C$ in Eq.~(\ref{c})
provides the next-nearest-neighbor coupling attached with a $\pi/2$ phase.
The $\sigma_z C$ term can be synthesized by periodically modulating
the frequencies of the cavities \cite{WangCaiLiuEtAl2016}. 

\begin{figure}
\includegraphics[width=0.95\columnwidth]{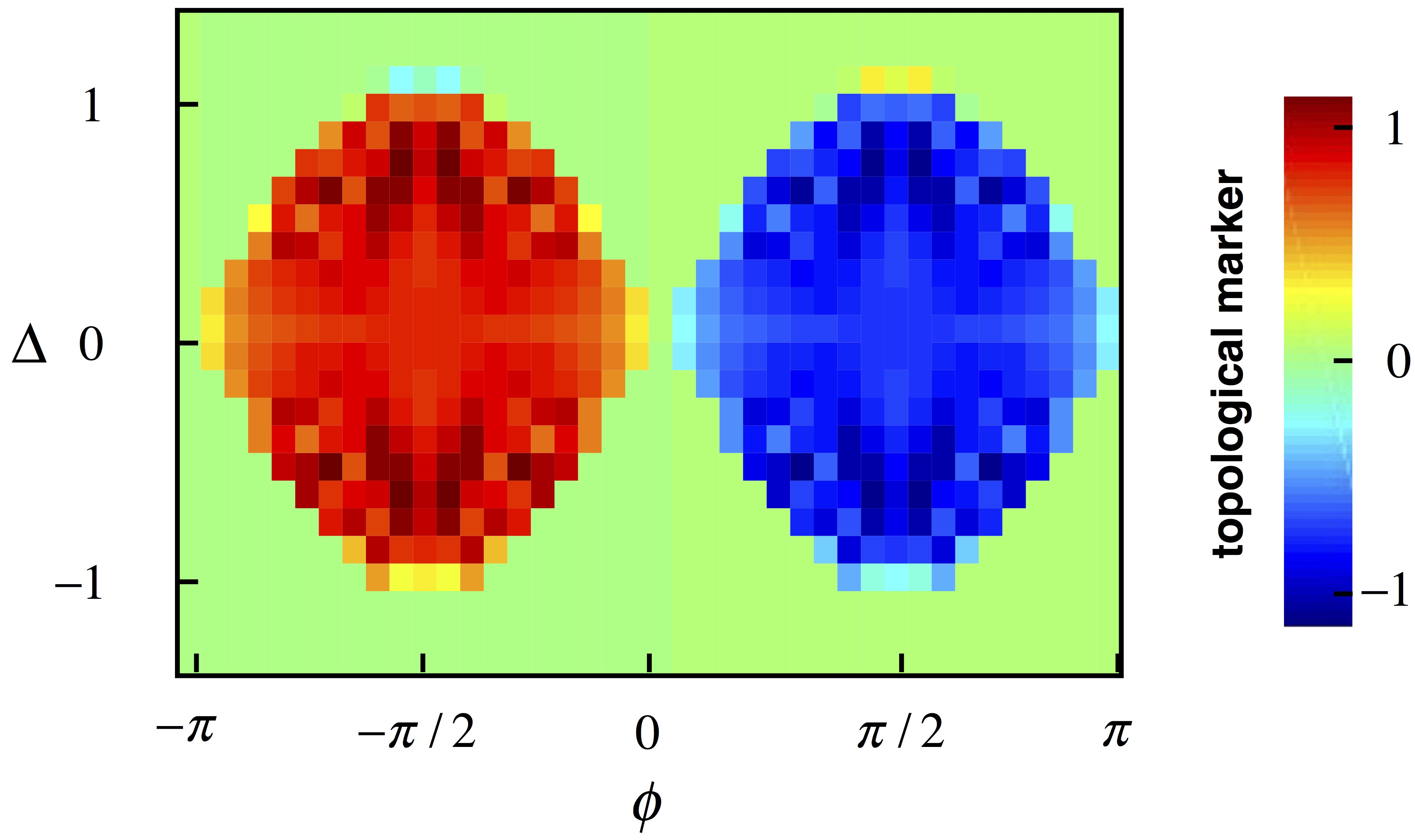}\caption{The topological marker of ${H}_5$ in Eq.~(\ref{h4}) evaluated in the center of the Fock-state lattice. $g=0.05$, $\kappa=2\sqrt{3}$, and $N=30$.}
\label{fig4}
\end{figure}

We plot the band structure of Eq.~(\ref{h4}) in Fig.~\ref{fig3} (b). The bulk states in 
the conduction and valence bands are generated from the eigenstates $|\psi^\pm_{m,C}\rangle$ in the Landau levels with $m\neq 0$, and their eigenenergies are $E^{\pm}_{m,C}=\pm \sqrt{mg^2+\kappa^2 C^2/4}$.
The eigenstates in the zeroth Landau level turn into the chiral edge states with eigenenergies $E_{0,C}=-\kappa C/2$ connecting the $K$ and $K^\prime$ points of the two bands.
The non-trivial topological property is
demonstrated by the unidirectional propagation of a wave packet of the edge states \cite{Rechtsman2013,Hafezi2013},
$\ket{\psi(0)} =(b_{+}^{\dagger}-b_{-}^{\dagger})^{N}{\ket{\downarrow,0,0,0}_b}/{\sqrt{2^{N}N!}}=i^N(a_{2}^{\dagger}-a_{3}^{\dagger})^{N}{\ket{\downarrow,0,0,0}}/{\sqrt{2^{N}N!}}$
which has zero mean energy.
With the weight located on the incircle (the boundary between the band insulator and the semimetal), the wave packet rotates clockwise (as shown in Fig.~\ref{fig3} (c)), which indicates the negative dispersion of the edge states in Fig.~\ref{fig3} (b).

In the original Haldane model \cite{Haldane1988}, the phase $\phi$ attached to the next-nearest-neighbor hopping can have values
different from $\pi/2$ and there is an energy off-set $\Delta$ between the two sublattices. 
A topological phase diagram can be plotted with respect to $\phi$ and $\Delta$.
The corresponding Hamiltonian in the FSL is,
\begin{equation}
\begin{aligned}
{H}_{5} = H_2+\frac{N\Delta}{2}\sigma_{z}+[\frac{\kappa}{2\sqrt{3}}e^{i\phi\sigma_z}\sum_{j=1}^{2}a_{j+1}^{\dagger}a_{j}+h.c.],
\label{h5}
\end{aligned}
\end{equation}
where $\Delta$ is the detuning between the frequencies of the cavities and that of the atom.
The Chern numbers are traditionally obtained in the reciprocal
space of lattices via Bloch wavefunction in a closed Brillouin zone \cite{thouless1982quantized}. Since the FSL
is finite with boundaries and non-uniform coupling strengths, the standard
way to obtain the Chern number is not applicable. Instead, the Chern numbers of $H_5$ are obtained through the local topological marker \cite{kitaev2006anyons,bianco2011mapping} in the center of the FSL (see Methods). They are plotted as a function of $\Delta$
and $\phi$ in Fig.~\ref{fig4}, which demonstrates the same topological phase diagram
as the original Haldane model \cite{Haldane1988}.

~\\ \textbf{Topological quantum responses with coherent light field.} The physics of topological quantum optics in the previous parts of the paper is based on the calculation with quantized Fock states. A natural question is whether some of these phenomena have classical correspondence and whether the topological properties can be observed with classical light. In particular, well-known classical phenomena of atom-light interactions shall be explained with the FSL.
For instance, if the classical fields of the three cavity modes have the same frequency and amplitude, and their phases are arranged in such a way that the field strengths cancel at the position of the atom, the atom shall be decoupled with the cavities. In the following, we show that this classical scenario corresponds to the decoupling of the atom and photons in the zeroth Landau level of the FSL. Then we make transition to the intrinsic topological quantum phenomena that can be demonstrated by classical light field but without interpretation in classical optics.

\begin{figure}

\includegraphics[width=0.95\columnwidth]{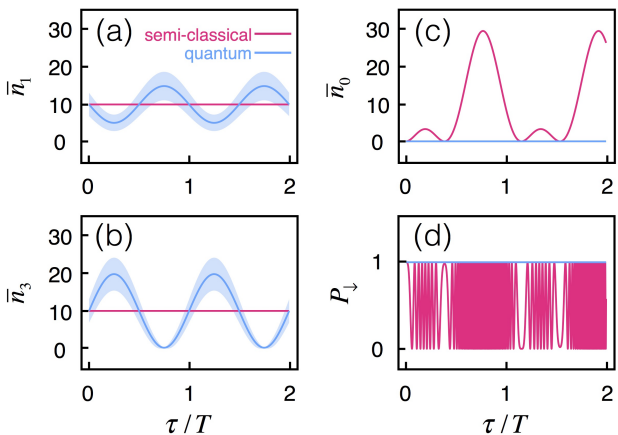}\caption{Evolution of the coherent cavity fields and atom with full quantum (blue lines) versus semiclassical (red lines) approaches. (a) The evolution of the average photon number $\bar{n}_1=\langle a_1^\dagger a_1\rangle$ (the same as that of mode $a_2$). (b) The evolution of $\bar{n}_3=\langle a_3^\dagger a_3\rangle$. (c) The evolution of the photon number in the bright mode $\bar{n}_0=\langle b_0^\dagger b_0\rangle$, which is coupled with the atom. (d) The evolution of the ground-state population of the atom. The blue shaded areas in (a) and (b) show the uncertainties of the cavity photon numbers $\Delta n_1$ and $\Delta n_3$ in the quantum approach. $g=1$, $\delta=0.1$, $|\alpha|^2=10$. In addition to the quantum fluctuations, the quantum approach demonstrates nontrivial dynamic evolution of the cavity fields with the atom remaining in the ground state, while the classical approach suggests Rabi oscillations of the atom and the photon numbers in the three cavities remain constant.}
\label{fig7}
\end{figure}

The state of the atom interacting with three classical light fields can be written as $\ket{\downarrow,\alpha_1, \alpha_2, \alpha_3}$, where $\ket{\alpha_j}=\exp({-|\alpha_j|^2/2)\sum_{n_j} \alpha_j^{n_j}\ket{n_j}}/\sqrt{n_j!}$ with $\alpha_j$ being complex numbers are the coherent states of the cavity modes $a_j$. We further assume $\alpha_j=\alpha \exp(-i2j\pi/3)$ such that the three fields cancel, i.e., $b_0\ket{\downarrow,\alpha_1, \alpha_2, \alpha_3}=0$. It is easy to understand classically that the atom is decoupled from the fields since it feels zero field strength. To explain this from the quantum perspective, we expand the state as a superposition of the eigenstates in the zeroth Landau levels of different subspaces (for $|\alpha|^2\gg 1$), 
\begin{equation}
\ket{\downarrow,\alpha_1, \alpha_2, \alpha_3}=\sum_N \frac{e^{\frac{N-N_0}{2}}}{(2\pi N)^{1/4}}\left(\frac{N_0}{N}\right)^{\frac{N}{2}}  \ket{\psi_{0,-N}^{(N)}},
\label{coherent}
\end{equation}
where $N_0=3|\alpha|^2$ is the total average photon number in the three modes, and $|\psi_{0,C}^{(N)}\rangle$ are the eigenstates in the zeroth Landau level of the subspace with total excitation number $N$. Here only the states at the $K^\prime$ point appear and $C=-N$. The probability of obtaining $|\psi_{0,-N}^{(N)}\rangle$ in Eq.~(\ref{coherent}) is approximately $(2\pi N)^{-1/2}\exp{[-(N-N_0)^2/2N]}$, i.e., following a Gaussian distribution centered at $N_0$. The result in Eq.~(\ref{coherent}) is remarkable since it indicates that even with coherent fields in the three cavities, we can prepare a state in the zeroth Landau level at the $K^\prime$ point, although in a superposition of states from different subspaces. Since the Hamiltonian conserves $N$, the evolution of the state $\ket{\downarrow,\alpha_1, \alpha_2, \alpha_3}$ can be treated separately in each subspace.

Starting from the state in Eq.~(\ref{coherent}), we show the difference between the predictions of semiclassical and full quantum approaches with the valley Hall effect in Fig.~\ref{fig7}. A detuning $\delta$ (see Eq.~(\ref{h3})) between the frequencies of modes $a_1$ and $a_2$ removes the cancelation of the three fields during the dynamical evolution. From the point of view in classical optics, the atom feels a finite field strength and will be excited (see Fig.~\ref{fig7} (d)). This excitation, however, is absent in the quantum treatment for $\delta\ll g$. According to the calculation in Methods, the state remains in the zeroth Landau level and the atom stays in the ground state. This intraband evolution is protected by the band gap $g$, which is also the vacuum Rabi splitting and the cyclotron frequency in the pseudomagnetic field. 

On the other hand, although the atom remains in $\ket{\downarrow}$, it has a substantial effect on the evolution of the cavity modes.
Without the atom, the three fields do not interact with each other and $\delta$ can only change the relative phases between them.
With the presence of the atom, the valley Hall effect induces exchange of photons between the three cavity modes such that the zero value of their superposition is maintained, as shown in Fig.~\ref{fig7} (a)-(c). The bright mode $b_0$ is a dynamical constant of the Hamiltonian in Eq.~(\ref{heff}) of Methods. Please also note that in order to keep $b_0$ being zero, a classical version of the relation in Eq.~(\ref{tb}), $||\alpha_1|-|\alpha_2||\leq |\alpha_3|\leq|\alpha_1|+|\alpha_2|$ must be satisfied, which is also consistent with the fact that the wavepacket is trapped within the incircle of the FSL (see Fig.~\ref{fig5}). For instance, at the time $\tau=3T/4$, the state in Eq.~(\ref{coherent}) evolves to $\ket{\downarrow, -\sqrt{6}\alpha e^{i\pi/6}/2, \sqrt{6} \alpha e^{i\pi/6}/2, 0}$, i.e., the cavity mode $a_3$ is in the vacuum state and the photons are equally distributed in modes $a_1$ and $a_2$. This is highly nontrivial since it demonstrates that the topological quantum phenomena discussed in this paper can be observed with the classical (coherent) field, but they cannot be explained without the second quantization of the light.
Similarly, the dynamics of the edge states of the Haldane model in Fig.~\ref{fig3} (c) can also be demonstrated with coherent light fields.\\

\noindent\textbf{Discussion}

\noindent In striking contrast to the photonic and acoustic topological insulators \cite{Lu2014,khanikaev2017two,ozawa2019topological,he2016acoustic}, where the topological properties do not require a quantization of the light field, all the topological properties discussed in this paper are based on the quantum nature of the bosonic operator,
i.e., $a\ket{n}=\sqrt{n}\ket{n-1}$ for $n\geq 1$ and $a\ket{0}=0$ (which ensures finite lattices with edges). 
Another difference from the photonic and acoustic topological insulator is that the FSL only needs a few modes to generate high dimensional lattices. $d$ bosonic modes can construct a FSL in $d-1$ dimensions, which offers a platform to simulate high-dimensional topological physics \cite{zhang2001four,lohse2018exploring,zilberberg2018photonic,Li2013,Li2012}. 

This study can also help to design novel artificial lattices for photons and phonons. A special type of lattices named the Glauber-Fock lattices \cite{perez2010glauber,keil2011classical} have been fabricated with waveguides, with the coupling strengths between neighboring waveguides mimicking the coupling between Fock states. These lattices can host collective modes that inherit the properties of the coherent state. In the same spirit, by replacing each state in the FSL with a cavity mode, we can construct a finite lattice of cavities that have a band structure similar to that in Fig.~{\ref{fig2}} (c), with each eigenstate being replaced by an eigenmode. Compared with the lattices designed with the strain-induced gauge field \cite{Guinea2010,Gomes2012,Yang2017,wen2019acoustic}, the lattice with coupling strengths between neighboring sites mimicking the 2D FSL has $\sqrt{m}$-scaling quantized energy levels everywhere, not limited near the $K$ and $K^\prime$ points.

The experimental realization of the physics discussed in the paper can be implemented in superconducting circuits with several resonators being coupled to a single qubit. In order to observe the dynamical process of the valley Hall effect and the chiral edge states of the Haldane model, we need the lifetime of the resonator $T_R$ satisfying $T_{R}/N\geq T,T_w$. Since only the zeroth Landau level with the qubit in the ground state is involved with these two phenomena, the decoherence from the qubit has no effect. For Landau-Zener tunneling, the atom can be in the excited state and thus it also requires $T_{a1}, T_{a2}\geq T,T_w$ where $T_{a1}$ and $T_{a2}$ are the lifetime and deocherence time of the qubit.
The state-of-the-art parameters are $T_{R}\approx 20\mu s$, $T_{a1}\approx 20\mu s$, $T_{a2}\approx 2\mu s$ and $g\approx 2\pi\times50\text{MHz}$ \cite{song2019generation} and $T_w\approx 450ns$ \cite{wang2019synthesis}. If we adopt a reasonable $T=200ns$ for $\delta=2\pi\times5\text{MHz}$, the above conditions can be satisfied with the excitation number $N=10$, which is sufficient to observe the topological phenomena.\\

\noindent\textbf{Acknowledgments}

\noindent We acknowledge that Ren-Bao Liu proposed \cite{Liu2010} the idea of using multi-mode Fock states of photons to simulate topological physics in lattices (such as the 2D Haldane model) and to simulate physics in higher-dimensional spaces (such as the four-dimensional quantum Hall effect). The authors would like to thank Zhaoju Yang, Congjun Wu and Jian Zi for insightful discussions.~This research is supported by National Key Research and Development Program of China (Grants No.~2019YFA0308100), the National Natural Science Foundation of China (No.~11874322), and the Fundamental Research Funds for the Central Universities.\\

\noindent\textbf{Author Contributions}

\noindent The two authors contribute substantially to all aspects of the manuscript.\\

\noindent\textbf{Competing Interests statement}

\noindent The authors declare no competing interests.\\

\noindent\textbf{Methods}

\noindent\textbf{Location of the soliton state in 1D Fock-state SSH model.} We expand $\left| \psi_s\right\rangle$ in the basis of $a$ modes,
\begin{equation}
\label{eq_soliton}
|\psi_s\rangle=\sqrt{\frac{N!}{n_1!n_2!}}u_2^{n_1}(-u_1)^{n_2}\ket{\downarrow,n_1,n_2},
\end{equation}
according to which Fig.~\ref{fig1} (b)-(d) are drawn. The probability distribution of the  dark state in Eq. (\ref{eq_soliton}) is
\begin{equation}
\begin{aligned}
|\langle \downarrow,n_1,n_2|\psi_s\rangle|^2\propto\frac{u_2^{2n_1}u_1^{2n_2}}{n_1!n_2!}.
\end{aligned}
\end{equation}
By using the Stirling's formula for $N\gg1$ and $1\ll n_1\ll N$, we obtain the condition of the distribution maxima in the lattice,
\begin{equation}
\begin{aligned}
\frac{\partial}{\partial n}\ln|\langle \downarrow,n_1,n_2|\psi_s\rangle|^2\end{aligned}\propto\ln{\left(\frac{u^2_2}{u_1^2}\frac{n_2}{n_1}\right)}=0,
\end{equation}
which results in
\begin{equation}
 \begin{aligned}
 u_1\sqrt{n_1} = u_2\sqrt{n_2},
 \end{aligned}
 \end{equation}
i.e., the state is centered at the point where the two neighboring coupling strengths are equal, and the photon number in $a_1$ mode is $n_1=u_2^2 N$.

~\\ \textbf{The eigen wavefunction in the zeroth Landau level.} Here we compare the wavefunction in the zeroth Landau level
near $K'$ point $|\psi_{0,-N}\rangle$ (see Fig.~\ref{fig2} (d)) with that in the Landau level of a real magnetic field in the symmetric
gauge, $\psi_{0LL}(r)\propto \exp({-r^{2}/4l_{B}^{2}})$ \citep{ezawa2008quantum}. We will show that $|\psi_{0,-N}\rangle$ is more localized in the Fock-state lattice than $\psi_{0LL}(r)$ due to the inhomogeneity of the magnetic field (see Fig.~\ref{fig2} (b) and Eq.~(\ref{Br})).
The probability distribution of $|\psi_{0,-N}\rangle$ is
\begin{equation}
|\langle\downarrow,n_{1},n_{2},n_{3}|\psi_{0,-N}\rangle|^2=\frac{N!}{3^{N}n_{1}!n_{2}!n_{3}!},
\end{equation}
which has rotational symmetry. Therefore, we
only need to consider the distribution along a radial direction from the center of the lattice to vertex 1 in Fig.~\ref{fig2} (a),
e.g., in states $\ket{p}\equiv |\downarrow,N/3-2p,N/3+p,N/3+p\rangle$ where $-N/3\leq p\leq N/6$
being an integer. By using the Stirling's formula for $N\gg 1$ and $p\ll N$, we obtain
\begin{equation}
|\langle p|\psi_{0,-N}\rangle|^2\propto  \exp\left({-\frac{9p^{2}}{N}}\right)=\exp\left(-\frac{2r^2}{l_B^2}\right),
\end{equation}
where the radius
$r$ is related to number $p$ through $r=3pq/2$. Comparing with $|\psi_{0LL}(r)|^2 \propto \exp({-r^{2}/2l_{B}^{2}})$, $|\psi_{0,-N}\rangle$ has a smaller variance.

\begin{figure*}[t]
\includegraphics[width=1.8\columnwidth]{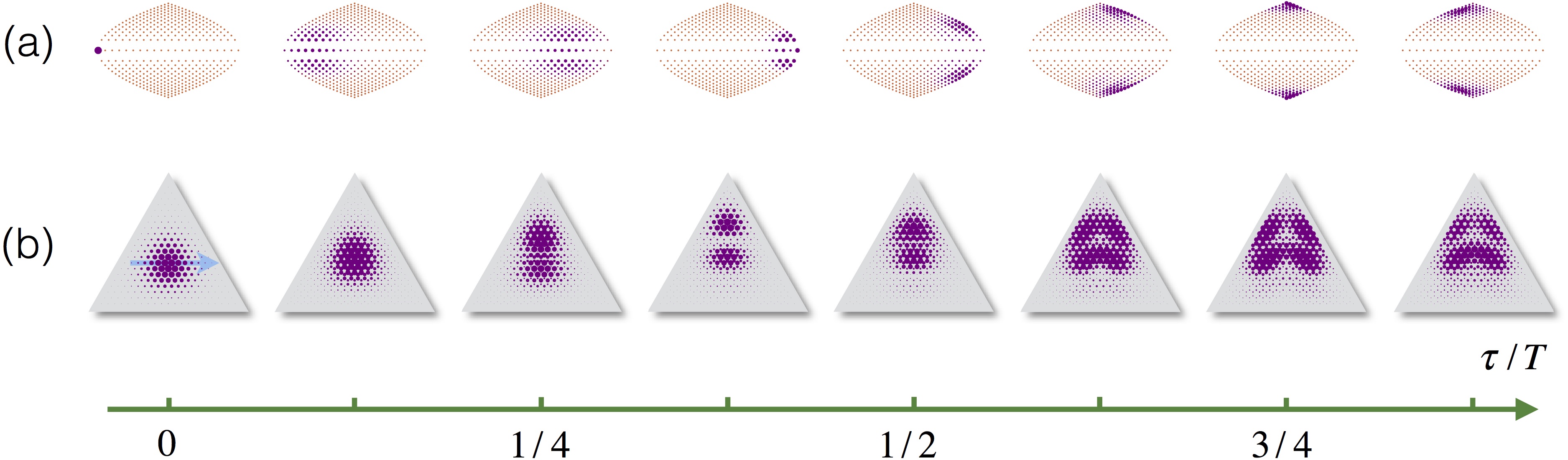}\caption{The Landau-Zener tunneling in the Fock-state lattice with $N=20$. (a) The Landau-Zener tunneling shown by the wavefunctions spreading to higher Landau levels due to a large effective force, $\delta=g$, in the direction shown by the blue arrow. (b) The evolution of the wavefunctions in the FSL. The radii of the purple solid circles are proportional to the probabilities in the corresponding states.}
\label{lzt}
\end{figure*}

~\\ \textbf{Bosonic chirality operator.} The spin chirality operator is defined for three spins, $C_s=\vec{\sigma}_1\cdot (\vec{\sigma}_2\times\vec{\sigma}_3)$ where $\vec{\sigma}_j$ is the Pauli matrix vector of the $j$th spin. 
To generalize the chirality operator to bosons, we need to notice that the spin chirality operator breaks both the parity $\mathscr{P}$ and time-reversal $\mathscr{T}$ symmetry, but conserves the $\mathscr{PT}$ symmetry \cite{Wen1989}. Another key feature of the spin chirality operator is that its evolution operator chirally rotates the spin states, $|s_1s_2s_3\rangle\rightarrow|s_2s_3s_1\rangle\rightarrow|s_3s_1s_2\rangle$ \cite{wang2019synthesis}. The chirality operator defined in Eq.~(\ref{c}) has these two properties. It is easy to verify that $\mathscr{P}C\mathscr{P}^{-1}=-C$, $\mathscr{T}C\mathscr{T}^{-1}=-C$ and $\mathscr{PT}C\mathscr{T}^{-1}\mathscr{P}^{-1}=C$. The evolution operator of $C$ rotates photons chirally among the three modes $a_1\rightarrow a_2\rightarrow a_3$ \cite{WangCaiLiuEtAl2016}.

~\\ \textbf{The valley Hall response.} In the limit of a small effective electric field, $\delta\ll g$, 
the evolution of an initial state in the zeroth Landau level is confined in that level.
We can study the evolution by projecting the Hamiltonian
$H_{3}$ in Eq.~(\ref{h3}) to the subspace of the zeroth Landau level,
\begin{equation}
\begin{aligned}
H_{\text{eff}} & ={P}_0H_{3}{P}_0,\\
 & =\frac{\delta}{\sqrt{3}}(e^{-i\pi/6}b_{+}^{\dagger}b_{-}+e^{i\pi/6}b_{-}^{\dagger}b_{+}),
\end{aligned}
\label{heff}
\end{equation}
where ${P}_0=\sum_{C}|\psi_{0,C}\rangle\langle\psi_{0,C}|$ is the
projection operator in the zeroth Landau level. The Heisenberg equations of the operators are
\begin{equation}
\begin{aligned}
\frac{d}{d\tau}b_{+}^{\dagger} & =i[H_{\text{eff}},b_{+}^{\dagger}]=-e^{-i\pi/3}\frac{\delta}{\sqrt{3}}b_{-}^{\dagger},\\
\frac{d}{d\tau}b_{-}^{\dagger} & =i[H_{\text{eff}},b_{-}^{\dagger}]=e^{i\pi/3}\frac{\delta}{\sqrt{3}}b_{+}^{\dagger}.
\end{aligned}
\end{equation}
We obtain the state evolution
\begin{equation}
\begin{aligned}
b_{+}^{\dagger}(0)=e^{-i\pi/3}b_{-}^{\dagger}(\tau)\sin\frac{\delta}{\sqrt{3}}\tau+b_{+}^{\dagger}(\tau)\cos\frac{\delta}{\sqrt{3}}\tau,\\
b_{-}^{\dagger}(0)=b_{-}^{\dagger}(\tau)\cos\frac{\delta}{\sqrt{3}}\tau-e^{i\pi/3}b_{+}^{\dagger}(\tau)\sin\frac{\delta}{\sqrt{3}}\tau.
\label{be}
\end{aligned}
\end{equation}
The evolution of the state $\left|\downarrow,0,0,N\right\rangle_{b}$ at $K'$ point is determined by
\begin{equation}
\frac{[b_{-}^{\dagger}(0)]^{N}}{\sqrt{N!}}|\downarrow,0,0,0\rangle_{b}  =\sum_{n}c_{n,N-n}(\tau)|\downarrow,0,n,N-n\rangle_{b},
\end{equation}
where $|c_{n,N-n}(\tau)|\propto|(\cos\frac{\delta}{\sqrt{3}}\tau)^{n}(\sin\frac{\delta}{\sqrt{3}}\tau)^{N-n}|$ can be obtained through Eq.~(\ref{be})
and the distribution is shown in Fig.~{\ref{fig5}} (a), demonstrating the Bloch oscillation in the zeroth Landau level,
\begin{equation}
|\downarrow,0,0,N\rangle_{b}\xrightarrow{T/2}|\downarrow,0,N,0\rangle_{b}\xrightarrow{T/2}|\downarrow,0,0,N\rangle_{b}.
\end{equation}
The $x$ and $y$ coordinates in the Fock-state
lattice are
\begin{equation}
\begin{aligned}
y& =\frac{q}{2}(2a_{3}^{\dagger}a_{3}-a_{1}^{\dagger}a_{1}-a_{2}^{\dagger}a_{2}),\\
x& =\frac{\sqrt{3} q}{2} (a_{2}^{\dagger}a_{2}-a_{1}^{\dagger}a_{1}).
\end{aligned}
\label{xy}
\end{equation}
Since $x$ commutes with $H_{\text{eff}}$, it does not change with time. This is a signature of the Hall response considering that the force is along the $x$ direction. Using Eq.~(\ref{xy}) and the inverse relation of Eq.~(\ref{be}) and considering the initial state $\left|\downarrow,0,0,N\right\rangle_{b}$, we obtain the evolution in the $y$ direction,
\begin{equation}
\begin{aligned}
y(\tau) =&- \frac{q}{2}\langle e^{-i\pi/3}b_-^\dagger(\tau)b_+(\tau)+e^{i\pi/3}b_+^\dagger(\tau)b_-(\tau)\rangle \\
=& \frac{q}{2} \langle b_-^\dagger (0) b_- (0)- b_+^\dagger (0) b_+ (0)\rangle\sin\frac{2\delta\tau}{\sqrt{3}}\\
=&R\sin\frac{2\pi\tau}{T}.
\end{aligned}
\end{equation}

\begin{figure}
\includegraphics[width=0.9\columnwidth]{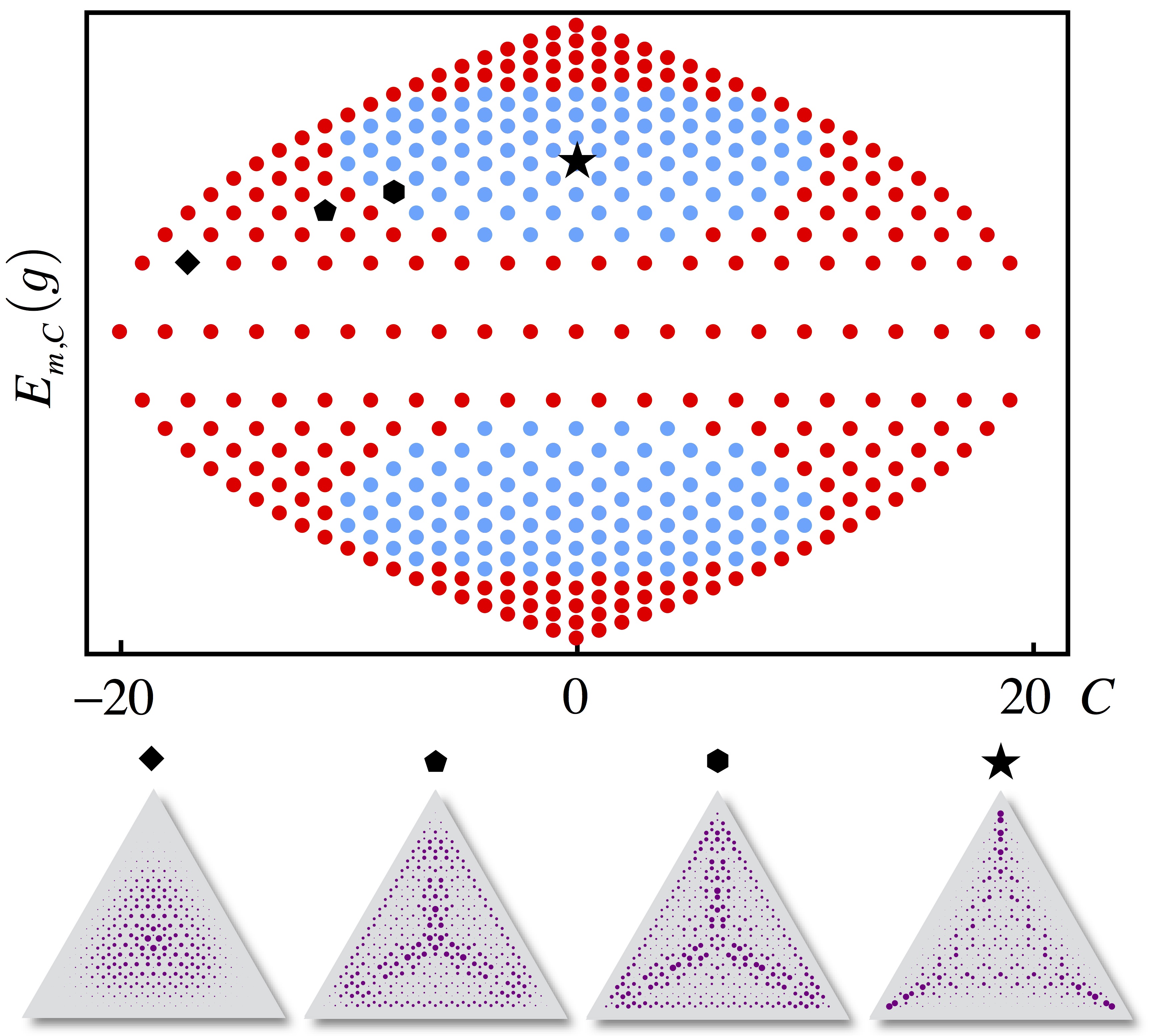}\caption{The eigenstates in (red) and out of (blue) the incircle. Four typical wavefunctions in the first (diamond), third (pentagon), fourth (hexagon) and sixth (star) Landau levels are plotted in the FSL.}
\label{wd}
\end{figure}

~\\ \textbf{The Landau-Zener tunneling and the wavefunctions beyond the zeroth Landau level.} The Landau-Zener tunneling appears when the potential difference between neighboring lattice sites $\delta$ is comparable with the bandgap $g$, as shown in Fig.~\ref{lzt}. While oscillating between the $K$ and $K^\prime$ points, the state tunnels to other Landau levels. In the FSL, the Landau-Zener tunneling add new features such as the splitting of the wavepackets into multiple components, demonstrating interference between states in different Landau levels.

The eigenstates in the zeroth Landau level of the 2D FSL is a two-dimensional extension of the soliton states in the 1D FSL. In particular, the weight of a wavepacket centered at the state $\ket{\psi_{0,0}}$ locates on the incircle, as shown by Fig.~4. Therefore, we can use $\ket{\psi_{0,0}}$ to set the boundary between the eigenstates in the strained semimetal and the band insulator. In Fig.~\ref{wd}, the eigenstates with variations $\langle r^2\rangle$ smaller than that of $\ket{\psi_{0,0}}$ are plotted red, otherwise blue, which can be regarded as in two different topological phases. 
Several typical eigen wavefunctions are also plotted. The one in the blue area occupies more sites outside of the incircle.

~\\ \textbf{The topological phase transition between a semimetal and a band insulator.} It has been shown that the strain can shift the Dirac cones of graphene, which has the effect of a vector potential until the anisotropy of
the strain is large enough to merge two Dirac cones into one, beyond which a bandgap opens \cite{pereira2009tight}. Here we show that the topological phase transition happens at the incircle of the triangular boundary of the FSL.
Considering the lattice site $\ket{\downarrow, n_1, n_2, n_3}$, the coupling strengths are $t_j=\sqrt{n_j}g/\sqrt{3}$. The condition for the semimetal phase in Eq.~(\ref{tb}) can be rewritten as,
\begin{equation}
n_1+n_2-2\sqrt{n_1n_2}<n_3<n_1+n_2+2\sqrt{n_1n_2}.
\label{tpt}
\end{equation}
From Eq.~(\ref{xy}) and the constrain $\sum_j n_j=N$, we obtain
\begin{equation}
\begin{aligned}
n_1&=\frac{Nq-y}{3q}-\frac{x}{\sqrt{3}q},\\
n_2&=\frac{Nq-y}{3q}+\frac{x}{\sqrt{3}q},\\
n_3&=\frac{N}{3}+\frac{2y}{3q}.
\end{aligned}
\label{n123}
\end{equation}
Substituting Eq.~(\ref{n123}) in Eq.~(\ref{tpt}), we obtain
\begin{equation}
x^2+y^2<R^2,
\label{inner}
\end{equation}
i.e., the sites are in the incircle of the triangular boundary. Substituting Eq.~({\ref{xy}}) in Eq.~(\ref{inner}), we obtain the relation of the photon numbers in Eq.~(\ref{innernumbers}).

~\\ \textbf{Pseudomagnetic field distribution obtained from the shift of Dirac
points.} At the Dirac points of a tight-binding honeycomb lattice, the Bloch wavevectors $\mathbf{k}$ satisfy the relation,
\begin{equation}
\left|t_{3}+t_{1}e^{-i\mathbf{k}\cdot\mathbf{v}_{1}}+t_{2}e^{-i\mathbf{k}\cdot\mathbf{v}_{2}}\right|=0,
\end{equation}
where $\mathbf{v}_{1}=(-\sqrt{3}q/2,-3q/2)$ and $\mathbf{v}_{2}=(\sqrt{3}q/2,-3q/2)$.
Accordingly, the positions of the Dirac points are explicitly obtained through the equations,
\begin{equation}
\begin{aligned}
\cos\mathbf{k}\cdot\mathbf{v}_{1}&=\frac{t_{2}^{2}-t_{1}^{2}-t_{3}^{2}}{2t_{1}t_{3}}\equiv s_{1},\\
\cos\mathbf{k}\cdot\mathbf{v}_{2}&=\frac{t_{1}^{2}-t_{2}^{2}-t_{3}^{2}}{2t_{2}t_{3}}\equiv s_{2}.
\end{aligned}
\label{v12}
\end{equation}
In the FSL the coupling strengths vary locally and the Dirac points shift at different locations. At the site $\ket{\downarrow,n_1,n_2,n_3}$, we obtain
\begin{equation}
\begin{aligned}
s_1&=\frac{n_{2}-n_{1}-n_{3}}{2\sqrt{n_{1}n_{3}}},\\
s_2&=\frac{n_{1}-n_{2}-n_{3}}{2\sqrt{n_{2}n_{3}}}.
\end{aligned}
\label{u12}
\end{equation}
From Eq.~(\ref{v12}), we obtain,
\begin{equation}
\begin{aligned}
k^\pm_{x}=&\pm\frac{1}{\sqrt{3}q}(\arccos s_{1}+\arccos s_{2}),\\
k^\pm_{y}=&\pm\frac{1}{3q}(\arccos s_{1}-\arccos s_{2}),
\end{aligned}
\label{kxy}
\end{equation}
where $\mathbf{k}^\pm=(k^\pm_x, k^\pm_y)$ with $+$ and $-$ denoting the two Dirac points $K$ and $K^\prime$. The shift of the Dirac points in the Brillouin zone is equivalent to a pseudo vector potential,
\begin{equation}
\mathbf{A}^\pm=(A_{x}^{\pm},A_{y}^{\pm})=\frac{\hbar}{e}(k^\pm_{x},k^\pm_{y}),
\end{equation}
which results in the pseudomagnetic field,
\begin{equation}
\begin{aligned}
B^{\pm}=\frac{\partial A_{y}^{\pm}}{\partial x}-\frac{\partial A_{x}^{\pm}}{\partial y}=\frac{\hbar}{e}\left(\frac{\partial k_{y}^{\pm}}{\partial x}-\frac{\partial k_{x}^{\pm}}{\partial y}\right).
\end{aligned}
\label{bst}
\end{equation}
Substituting Eqs.~(\ref{n123}), (\ref{u12}) and (\ref{kxy}) in Eq.~(\ref{bst}) and after a cumbersome algebraic calculation, we obtain,
\begin{equation}
\begin{aligned}
B^{\pm} & =\mp\frac{2\hbar}{Neq^{2}}\text{\ensuremath{\frac{1}{\sqrt{1-4r^{2}/q^2N^{2}}}}}\\
 & =\mp\frac{B_{0}}{\sqrt{1-r^{2}/R^{2}}},
\end{aligned}
\end{equation}
which is consistent with the one obtained from the valley Hall effect in Eq.~(\ref{Br}).

\begin{figure}
\includegraphics[width=0.8\columnwidth]{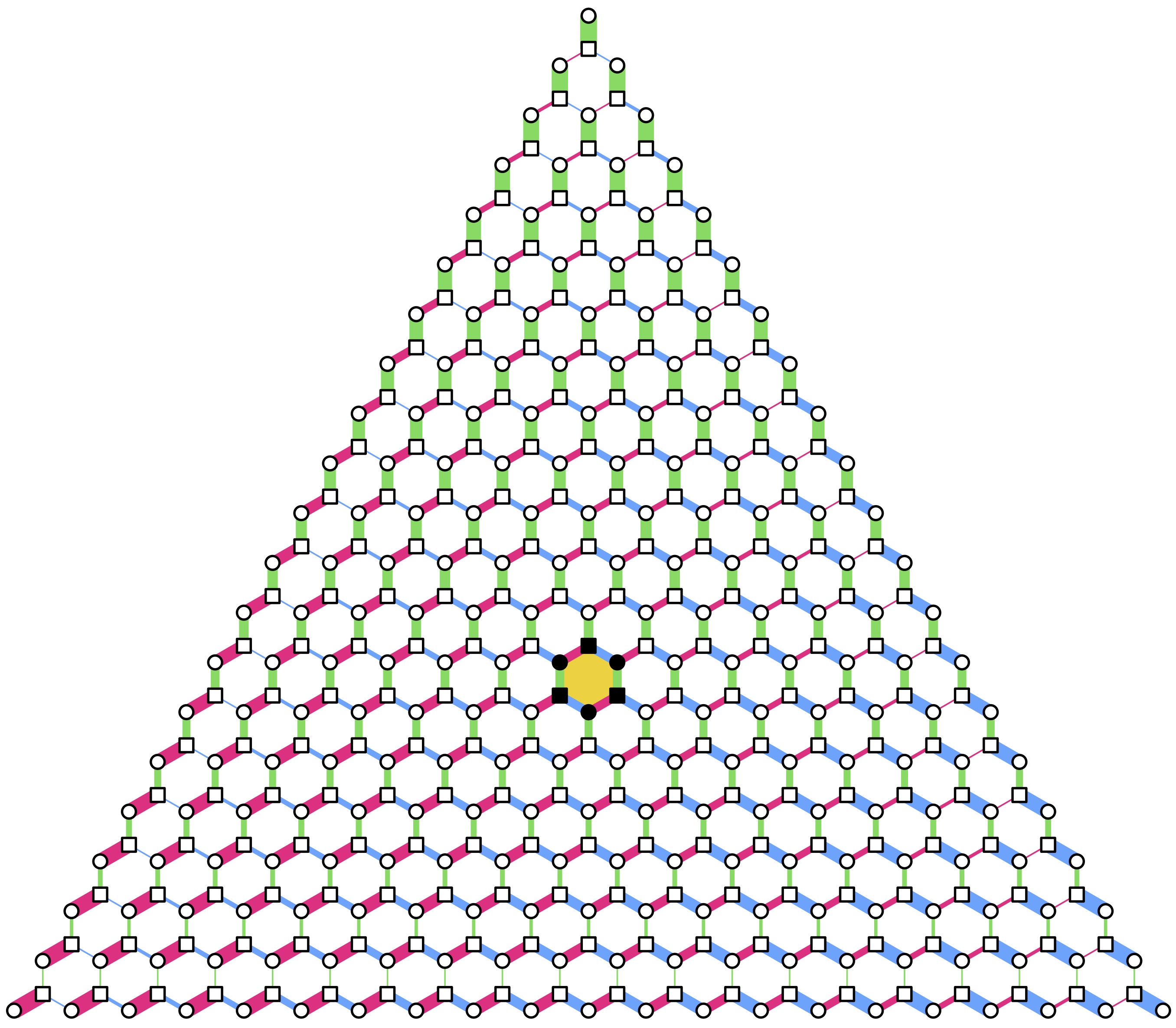}
\caption{The Fock-state lattice (with $N=20$) where the topological marker is calculated in the colored unit cell with the six vertices filled with black color.}
\label{figs2}
\end{figure}

~\\ \textbf{Topological marker.} We calculate the local topological marker in the filled
hexagon in Fig. \ref{figs2} according to ref.~\citep{bianco2011mapping},
\begin{equation}
M=\frac{8\pi}{9\sqrt{3}q^2}\text{Im}\sum_{s\in \varhexagon}\bra s {P}_- x {P}_- y{P}_-\ket s
\end{equation}
where $\ket s$ are the Fock states in the unit
cell colored in Fig.~\ref{figs2}, ${P}_-=\sum_{m,C}|\psi_{m,C}^{-}\rangle\langle\psi_{m,C}^{-}|$ is the projection operator of the lower band  where the eigenstate 
\begin{equation}
\begin{aligned}
|\psi^{-}_{m,C}\rangle=&-\sin\frac{\theta}{2}\left|\downarrow,m,m_+,m_-\right\rangle_b\\
&+\cos\frac{\theta}{2}\left|\uparrow,m-1,m_+,m_-\right\rangle_b,
\end{aligned}
\end{equation} 
with $\tan \theta=\kappa C/2g\sqrt{m}$, $x$ and $y$ are position operators in Eq.~(\ref{xy}). The topological marker $M$ is plotted in Fig.~5 for the phase diagram of the FSL Haldane model.

\end{document}